\documentclass[a4paper, 12pt]{article}

\usepackage[utf8]{inputenc}
\usepackage[T1]{fontenc}
\usepackage{amsmath}
\usepackage{psfrag}
\usepackage[cmintegrals]{newtxmath}
\usepackage{bm}
\usepackage{graphicx}
\usepackage{gensymb}
\usepackage[margin=1in]{geometry}
\usepackage{newfloat}
\DeclareFloatingEnvironment[name={Supplementary Figure}]{suppfigure}

\pdfoutput=1

\title{Fast and selective super-resolution ultrasound \textit{in vivo} with sono-switchable nanodroplets}

\author{
Kai Riemer$^1$, Matthieu Toulemonde$^1$, Jipeng Yan$^1$,\\
Marcelo Lerendegui$^1$, Eleanor Stride$^2$, Peter D. Weinberg$^1$, \\ Christopher Dunsby$^3$, Meng-Xing Tang$^1$*}

\date{%
    \footnotesize
    $^1$Dept. of Bioengineering, Imperial College London, SW7 2AZ, London, United Kingdom\\%
    $^2$NDORMS, University of Oxford, OX3 7HE, Oxford, United Kingdom\\%
    $^3$Dept. of Physics, Imperial College London, SW7 2AZ, London, United Kingdom\\%
    **Corresponding author: Meng-Xing Tang (Mengxing.tang@ic.ac.uk) \\%
}

\begin{document}

\maketitle

\begin{abstract}
Perfusion by the microcirculation is key to the development, maintenance and pathology of tissue. Its measurement with high spatiotemporal resolution is consequently valuable but remains a challenge in deep tissue. Ultrasound Localization Microscopy (ULM) provides very high spatiotemporal resolution but the use of microbubbles requires low contrast agent concentrations, a long acquisition time, and gives little control over the spatial and temporal distribution of the bubbles. The present study is the first to demonstrate Acoustic Wave Sparsely-Activated Localization Microscopy (AWSALM) and fast-AWSALM for \textit{in vivo} super-resolution ultrasound imaging, offering contrast on demand and vascular selectivity. Three different formulations of sono-switchable contrast agents were tested. We demonstrate their use with ultrasound mechanical indices well within recommended safety limits to enable fast on-demand sparse switching at very high agent concentrations. We produce super-localization maps of the rabbit renal vasculature with acquisition times between 5.5 s and 0.25 s, and an 4-fold improvement in spatial resolution. We present the unique selectivity of AWSALM in visualizing specific vascular branches and downstream microvasculature, and we show super-localized kidney structures in systole and diastole with fast-AWSALM. In conclusion we demonstrate the feasibility of fast and selective measurement of microvascular dynamics \textit{in vivo} with subwavelength resolution using ultrasound and sono-switchable nanodroplets.
\end{abstract}


\section{Introduction}
The microvasculature plays a critical role in the functioning of healthy tissue and changes in its structure and dynamics are an important diagnostic indicator in many diseases. For example, angiogenesis regulates the rate of tumor growth \cite{Lugano2020,Carmeliet2000} while deficits in myocardial perfusion by the microcirculation can predict adverse cardiac outcome \cite{Betancur2018}. Although the implications of structural anomalies in the microvasculature are relatively well understood, the impact of changes in real-time perfusion is less well known. In part this is due to the low spatial and/or temporal resolution of existing imaging modalities \cite{Dewey2020, Kierski2020} and the persistence of conventional contrast agents, which allow assessment of perfusion only through kinetic modelling \cite{Dewey2020}. A tool for the measurement of vascular structures with high spatiotemporal resolution and selective spatiotemporal activation would consequently be valuable for understanding dynamic processes in the microcirculation.

Super-resolution (SR) through optical switching and localization of individual fluorophores has revolutionized optical microscopy by enabling visualization of structures far below the wave diffraction limit \cite{Betzig2006, Rust2006}. Similarly, in acoustics, ultrasound localization microscopy (ULM) has enabled imaging of the microvasculature below the diffraction limit to create functional SR images of microvascular structures \cite{Christensen-Jeffries2015, Errico2015a}. Microbubble-based SR ultrasound techniques have shown unprecedented levels of detail of the microcirculation in pre-clinical imaging of the kidney \cite{Andersen2020, Chen2020}, lymph node \cite{Zhu2019b} and brain \cite{Hingot2017}, while clinical applications include breast \cite{Dencks2019}, lower limb \cite{Harput2018} and transcranial brain imaging \cite{Demene2021}. However, microbubble-based SR imaging requires seconds to minute-long acquisition times and cannot use a high bubble concentration without degrading the SR image quality as then individual bubbles cannot be localized. Some statistical approaches have been applied to permit higher concentrations and faster scan times to be used \cite{Bar-Zion2017a, VanSloun2018, Yu2018, Huang2020, Montreal2020}, but once a microbubble bolus has been injected, its concentration can only be regulated by deliberately destroying microbubbles with ultrasound or waiting for the bolus to become diluted by dispersion and partial clearance. It also takes time for smaller vessels to be adequately perfused with microbubbles due to the low flow rates \cite{Christensen-Jeffries2019, Hingot2019abc}.

These fundamental limitations arise because microbubble contrast agents are not switchable in the same way as fluorophores in optical SR microscopy. Phase-change ultrasound contrast agents ("nanodroplets"), on the other hand, are acoustically switchable and enable on-demand sparse activation and deactivation. In their initial state nanodroplets are not detected by ultrasound, acoustic activation leads to their detection and acoustic deactivation leads to disintegration. This can be achieved at very high agent concentrations, both with and without flow because not all nanodroplets are activated at the same time \cite{Zhang2018}. Nanodroplets are an order of magnitude smaller than microbubbles and can have a significantly longer half-life \textit{in vivo} \cite{Kee2019}. Low-boiling-point nanodroplets can be fabricated by condensation of fluorocarbon microbubbles \cite{Sheeran2011a}. While higher-boiling point droplets have been successfully triggered with a laser pulse \cite{Yoon2018}, low- boiling-point nanodroplets can be vaporized by the ultrasound pulse itself. Acoustic Wave Sparsely-Activated Localization Microscopy (AWSALM) \cite{Zhang2018} and fast-AWSALM \cite{Zhang2019a} using such acoustically switchable nanodroplets have shown flow independent and sub-second SR imaging \textit{in vitro} in a crossed tube phantom. An \textit{in vivo} demonstration of these techniques has, however, not yet been reported.

In this study we perform AWSALM and fast-AWSALM \textit{in vivo} to demonstrate fast and selective SR imaging using three different fluorocarbon nanodroplet compositions. We quantify the pressure dependent droplet activation in a crossed tube phantom and illustrate droplet vaporization. Finally, we show for first time sub-second super-localization images of diastolic and systolic perfusion \textit{in vivo} in the rabbit kidney and the selective imaging of specific vascular branches and downstream microvasculature.

\section{Materials and methods}

\subsection{AWSALM and fast-AWSALM}
\begin{figure}[h]
  \centering
    \includegraphics[width=8.9cm]{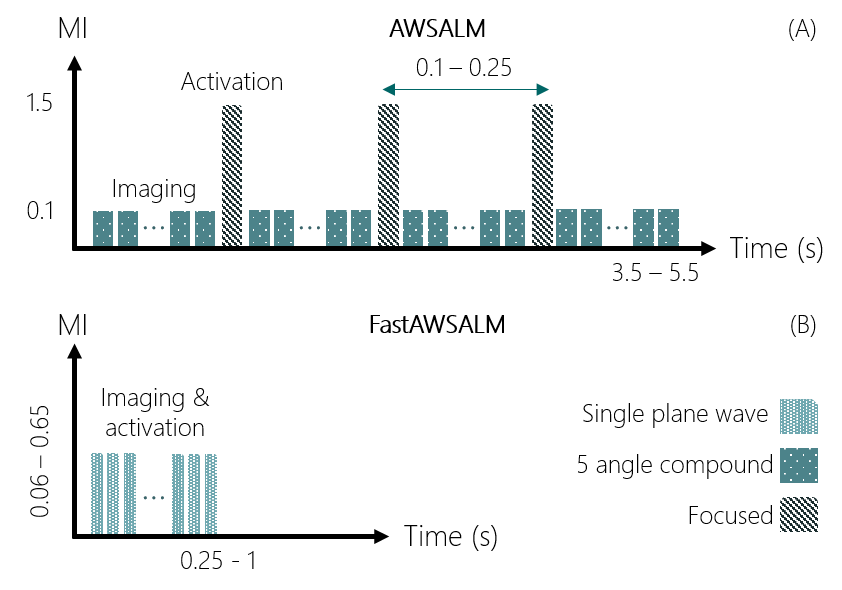}
      \caption[Activation sequence]{(A) The AWSALM sequence alternates between focused activation and 5 angle plane wave imaging. In this work, the number of focal points (1 or 2) and f-number were varied. The acquisition time was between 3.5-5.5 s with a frame rate of 500 Hz. (B) fast-AWSALM combines imaging and activation through a single plane wave with a maximum acoustic pressure located at a depth of 14 mm. In this work, the acquisition time was between 0.25-1 s and the MI was varied between 0.06-0.65, with frame rates up to 10,000 Hz}
  \label{fig:01}
\end{figure}

To sono-switch nanodroplets, both AWSALM and fast-AWSALM use acoustic transmissions that vaporize the liquid core of a condensed nanodroplet and lead to its subsequent expansion and formation of an oscillating microbubble. The two imaging techniques differ in their transmission sequence, acquisition duration and acoustic pressure. The latter is within recommended safety limits for diagnostic imaging. The AWSALM and fast-AWSALM pulse parameters used in this study are shown in Figure \ref{fig:01}.

AWSALM separates the transmitted pulse sequence used for imaging from the pulses used for selective activation and deactivation of droplets. Nanodroplets are repeatedly activated and imaged. With every focused transmission at a high mechanical index (MI), a sub-population of the nanodroplets within a predefined region are activated. These are then imaged by low MI plane wave imaging sequences followed by localization and tracking in post-processing. With each activation sequence new microbubbles are formed and existing bubbles are destroyed. This requires the use of long-lasting, medium-boiling point droplets \cite{Zhang2018}.

In fast-AWSALM, the same plane-wave transmission simultaneously sono-switches and images the contrast agent. With every transmission a sub-population of nanodroplets is vaporized. With every subsequent transmission a further sub-group of nanodroplets is vaporized and some of the previous bubbles are destroyed. The survival time of a vaporized bubble depends on the acoustic pressure and the frame rate. The use of metastable, low-boiling point droplets allows high concentrations to be used while still providing sparsity \cite{Zhang2019a}. The pressure required is lower than that for the focused transmission in AWSALM.

\subsection{Preparation of nanodroplet contrast agent}

The lower the boiling point of a perfluorocarbon, the lower is the energy required for nanodroplet activation and the higher is the probability of spontaneous vaporization. \textit{In vivo}, octafluoropropane (OFP) droplets are metastable and require very low acoustic pressures for vaporization, if any \cite{Mountford2015}. Decafluorobutane (DFB) droplets are more stable but therefore require higher acoustic pressures for activation. Through tuning, e.g. mixing gases or changing lipids, the metastable characteristics can be altered \cite{Mountford2014}.

For this study, three low-boiling-point fluorocarbon phase-change nanodroplets were made through condensation of phospholipid-coated microbubbles containing OFP (C3F8, b.p. -36.7\degree C), DFB (C4F10,  b.p. -1.9\degree C) or a mixture (C3F8:C4F10) (F2 Chemicals Ltd, UK). Microbubbles were fabricated as described by Lin et al. \cite{Lin2018}. Nanodroplets were formed from them as follows: a phospholipid suspension was prepared by mixing 1,2-distearoyl-sn-glycero-3-phosphocholine (DSPC) and 1,2-distearoyl-sn-glycero-3-phosphoethanolamine-N-[methoxy (polyethylene glycol) - 2000] (DSPE-PEG-2000) (9:1, molar ratio) in phosphate-buffered saline, propylene glycol, and glycerol (16:3:1 by volume) for a total lipid concentration of 1 mg/ml. One ml of the solution was added to a 2 ml glass vial. The head space was then filled with C3F8, C4F10 or a 1:1 volume ratio mixture. Microbubbles were formed through 90 s of mechanical agitation (TP-103 amalgamator, Goldsmith \& Reverse, Inc, USA). The vial containing microbubbles was subsequently submerged in a -10\degree C water-salt solution where it was kept 10 minutes until needed. To fabricate droplets, the cooled microbubble solution was condensed inside a 1 ml syringe by manually pushing the plunger with the end sealed. Droplets from a single vial were used for up to 30 minutes.

\subsection{Ultrasound scanner}
Experiments were conducted with a Verasonics Vantage 256 ultrasound research platform and a L11-4v linear array. The MI was calibrated in a water tank with a 0.2 mm needle hydrophone (Precision Acoustics, UK) and derated assuming a soft tissue attenuation coefficient of 0.5 dB/(MHz$\cdot$cm).
\subsection{Ultrafast imaging sequence}
Two different transmission sequences were used (Figure \ref{fig:01}). For AWSALM, the imaging part of the sequence consisted of 5 angle compounding single cycle plane waves at MI=0.1 (pulse centre frequency = 4 MHz), giving a compounded frame rate of 500 Hz for imaging. A twenty cycle focused transmission at MI=1.5 (centre frequency = 4 MHz) and frame rate of 50 Hz was used for vaporization. The number of focal depths was varied between 1 or 2, the focus depth was manually selected between 5-30 mm, the lateral focus position was selected between -15 to 15 mm. The number of activation/imaging cycles was between 18-25 and the total number of frames acquired ranged between 1,530 to 2,280. The total duration of an AWSALM acquisition was between 3.5-5.5 s; of which approximately 80\% was for imaging and 20\% for activation.

The fast-AWSALM sequence consisted of 3 cycle, single-angle, single-cycle plane wave imaging (centre frequency = 4 MHz) and frame rates between 2,000-10,000 Hz. The MI was varied between 0.06-0.65 with a peak MI at 14 mm depth. The total duration of each fast-AWSALM acquisition was between 0.25-1 s. A minimum interval of 20 s was maintained between fast-AWSALM acquisitions.

\subsection{Super-localization pipeline}
The radio frequency data were reconstructed with a delay-and-sum beamformer. A cross-correlation based rigid motion compensation and time gain compensation correction were applied prior to Singular Value Decomposition (SVD) clutter suppression. The AWSALM dataset was SVD filtered separately for each stack of images, whereas the fast-AWSALM dataset was filtered with all frames at once. Subsequently, the clutter-free data were filtered with an axial-temporal Wiener noise filter and a low-pass 3D convolutional filter. For the images of rabbit kidney acquired \textit{in vivo}, a region of interest was drawn manually around the renal vasculature for each acquisition. All subsequent steps of the localization are based on previous work \cite{Christensen-Jeffries2015}\cite{Zhang2018}. First, 20 vaporized droplets were manually selected to configure the system’s point spread function (PSF). Bubbles were localized in subsequent frames to create a velocity map based on cross-correlation with the PSF. Tracks were estimated and regularized using a combination of Kalman filter and graph-based assignment \cite{Yan2022}. For AWSALM, the interruptions to the image sequence by the focus transmissions were accounted for.

\subsection{\textit{In vitro} crossed tube}
Vaporization of the three different fluorocarbon nanodroplets was first assessed in a microvascular flow mimicking phantom with the low MI fast-AWSALM sequence. The setup consisted of two crossing cellulose tubes (Hemophan Membrana, 3M, Minnesota, USA) with an inner diameter of 200 $\pm$ 15 \textmu m. The crossed tube phantom was submerged in 37 $\pm$ 0.2 \degree C warm water inside a 20 L water tank. The volume flow was 30 \textmu l/min in each tube, resulting in a mean flow velocity of 16  mm/s. 0.5 ml of droplet solution was diluted in 15 ml of water. The centre of the crossed tubes was positioned at a depth of 15 mm with flow entering the field of view from the tube inlets furthest away from the transducer. The total acquisition time was 1 s with a frame rate of 5,000 Hz. Nanodroplet activation was quantified on unfiltered gray scale images by measuring the mean intensity of the tubes with a manually defined region of interest as a surrogate for vaporization activity. The intensity was normalized with respect to the MI and by the C4F10 acquisition at MI=0.06. Measurements were repeated 3 times. Supplementary Figure 1 shows the experimental setup and supplementary video 1 shows the droplet activation in the crossed tubes with AWSALM and fast-AWSALM. For visualization purposes only supplementary video 1 shows a very low droplet concentration.

\subsection{Renal imaging \textit{in vivo}}
Imaging was performed on the left kidney of anaesthetized specific-pathogen-free male New Zealand White rabbits. All experiments complied with the Animals (Scientific Procedures) Act 1986 and were approved by the Animal Welfare and Ethical Review Body of Imperial College London. Six rabbits (HSDIF strain, mean age 13 weeks; mean weight 2.4 kg; Envigo, UK) were sedated with acepromazine (0.5 mg/kg, i.m.) and anaesthetized with medetomidine (Domitor, 0.25 mL/kg, i.m.) plus ketamine (Narketan, 0.15 mL/kg, i.m.). The anaesthesia was maintained for up to 4 hours with administration of 1/3 of the initial medetomidine and ketamine dose every 30 minutes. Nanodroplet contrast agents were injected as a bolus of up to 0.3 ml/injection and a total of 1 ml/animal. Pentobarbital (0.8 ml/kg) was used for euthanasia. To access the renal vasculature, the rabbits were shaved and positioned supine. Mechanical ventilation was given at 40 breaths/minute; body temperature was maintained with a heated mat. Heart rate and oxygenation were continuously monitored.

\section{Results}

\subsection{\textit{In vitro} crossed tube evaluation}
\begin{figure}[ht!]
  \centering
    \includegraphics[width=14cm]{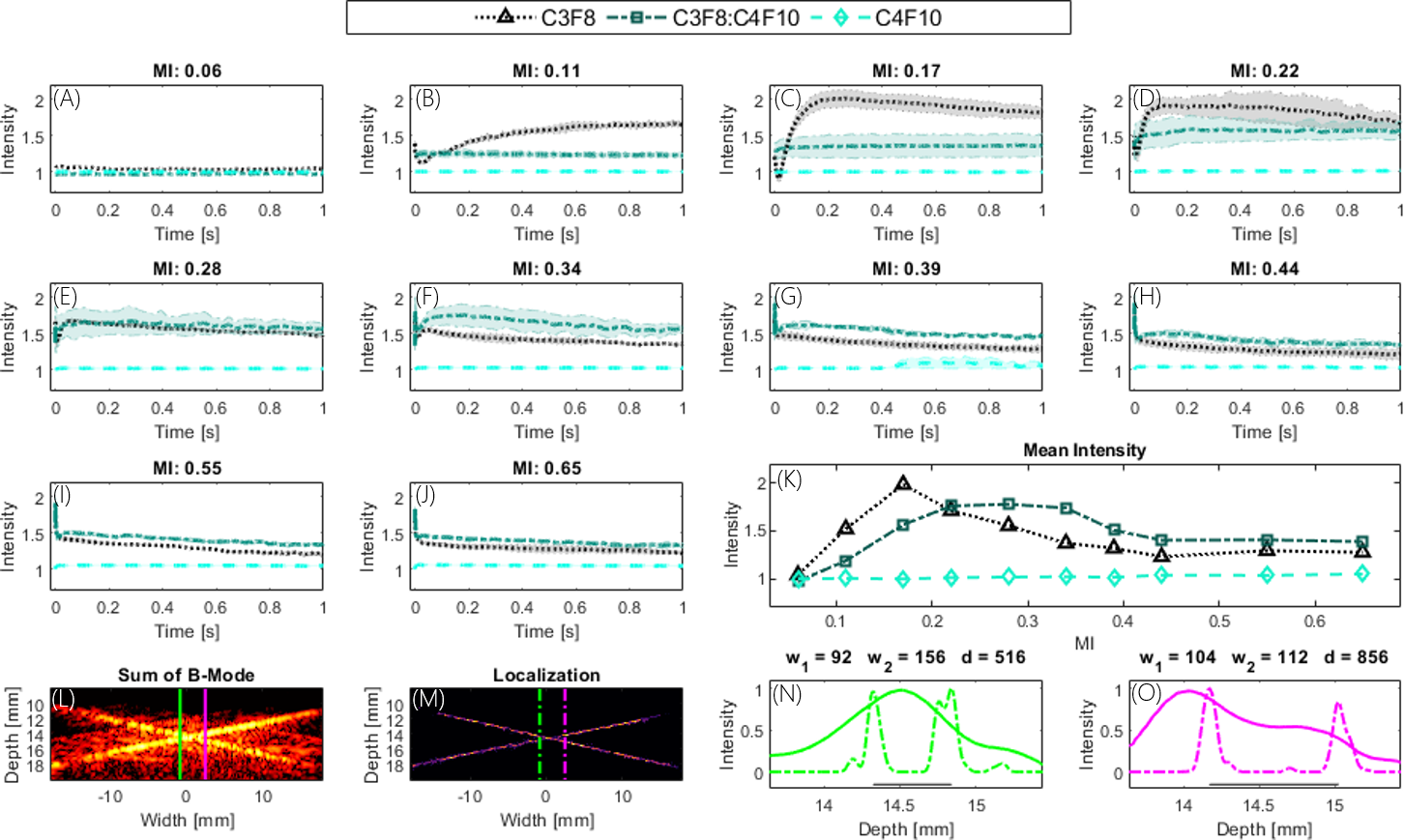}
      \caption{(A-J) mean intensity of the crossed tubes over time and (K) as a function of MI for different nanodroplet formulations. (L,M) sum of B-Mode and super-localization. (N,O) normalized intensity profile across green and magenta lines in (L,M). Both the C3F8 and mixed-gas droplets, but not the C4F10 droplets could be activated with plane waves at MIs between 0.17 and 0.65 (Figure \ref{fig:00} K). MIs>0.34 burst vaporized bubbles rapidly.}
  \label{fig:00}
\end{figure}

The changes in mean image intensity over time due to the vaporization of fluorocarbon nanodroplets (C3F8, C3F8:C4F10 and C4F10) in a crossed tube phantom with flow are shown in Figure \ref{fig:00} (A-J). For C3F8 and the mixture, respectively, the mean image intensity increased at MI=0.11/ MI=0.17 and peaked at MI=0.17/ MI=0.28 as shown in Figure \ref{fig:00} (K). At MI=0.06 nanodroplets did not vaporize, and C4F10 nanodroplets could not be vaporized with plane-waves at 37\degree C over the entire MI range tested. Repeated transmission at higher acoustic pressures caused a drop in intensity beyond MI=0.34, suggesting more destruction of the vaporized droplets at higher MIs. Figure \ref{fig:00} (L-O) show the images and normalized intensity profiles across two different lines of the sum of the B-Mode and super-localized results respectively. The crossed tubes can be resolved over a distance smaller than the axial direction of the PSF (see supplementary Figure 2).

\subsection{Vaporization, lifetime and destruction}
\begin{figure}[ht!]
  \centering
    \includegraphics[width=15cm]{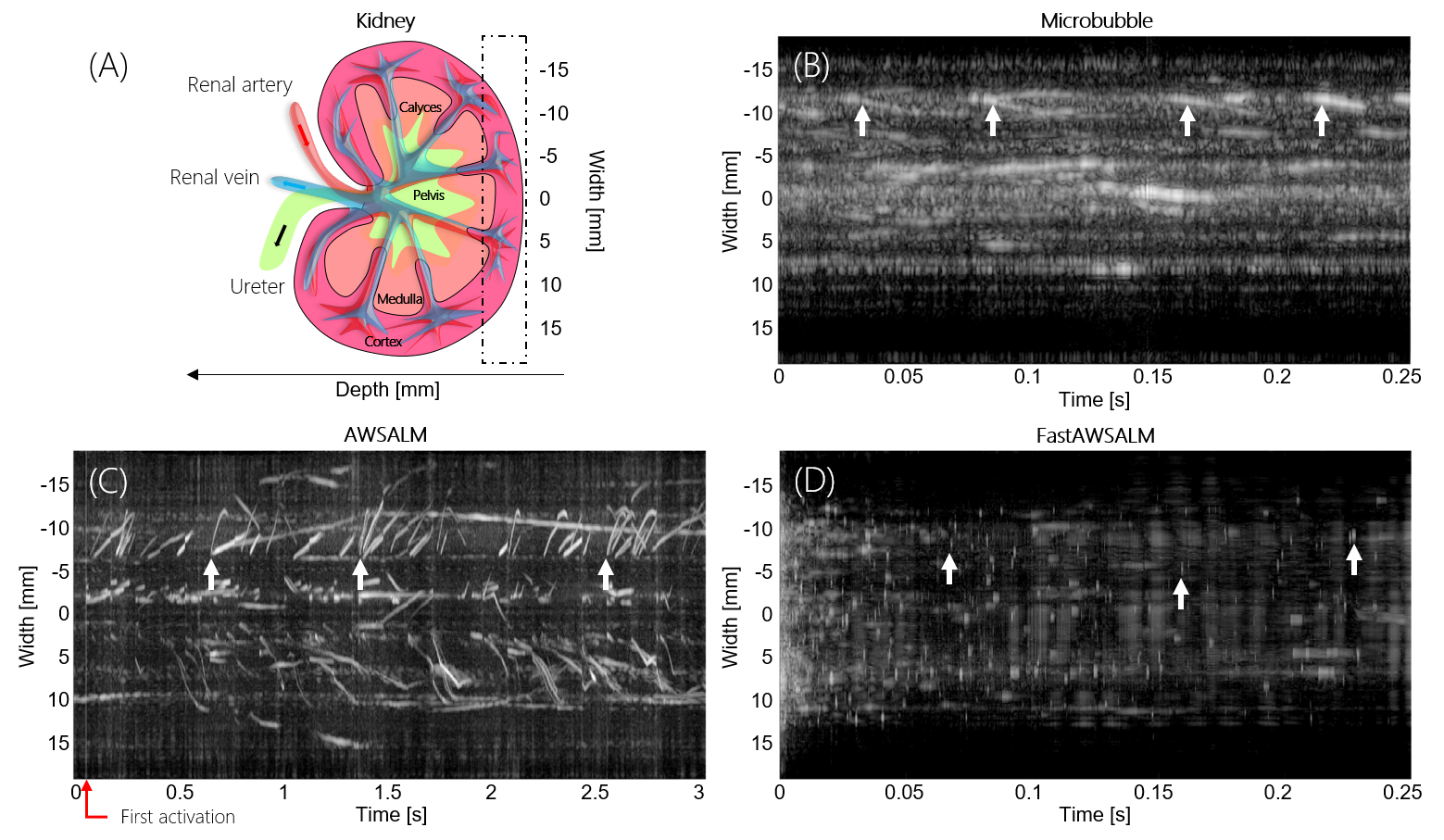}
      \caption{Spatiotemporal imaging of nanodroplet dynamics \textit{in vivo} in the rabbit kidney with conventional contrast, AWSALM and fast-AWSALM. The lateral-temporal signal of a maximum projection in the depth direction of the upper cortex of the rabbit kidney, as illustrated in (A, dashed box), obtained using (B) microbubbles, (C) AWSALM, (D) fast-AWSALM. Bubble tracks are visible in (B) and (C) as indicated by white arrows. Sparse activation and destruction of droplets can be observed in (D, white arrows). The lifetime from vaporization to destruction ranges from a few to dozens of transmissions. The first vaporization sequence of AWSALM is marked by the red arrow below (C). Note that Figure \ref{fig:02} (B/D) is from a single angle plane-wave acquisition of 0.25 s duration, whereas Figure \ref{fig:02} (C) is from a five-angle compounding acquisition of 3 s duration.}
  \label{fig:02}
\end{figure}

The difference between microbubble and nanodroplet spatio-temporal features and the influence of the imaging sequence in the rabbit kidney are shown in Figure \ref{fig:02}. Figure \ref{fig:02} (B-D) shows the lateral-temporal signal from a maximum projection taken in the depth direction of the upper cortex of the rabbit kidney (A, dashed box). Each acquisition was acquired after a bolus injection of 0.1 ml of contrast agent. The concentration of microbubbles in the maximum projection image is very high and only a few isolated bubble traces can be seen in parts of Figure \ref{fig:02} (B). Figure \ref{fig:02} (C) shows repeated activation of droplets with the AWSALM sequence. Droplets are sparse and the individual tracks are easily identifiable (examples shown by the white arrows). The red arrow below Figure \ref{fig:02} (C) marks the first activation event and the recurring activation of droplets led to a sparse, continuous stream of new bubbles. Figure \ref{fig:02} (D) presents the sparse activation and destruction of nanodroplets with fast-AWSALM, with exemplar events illustrated by white arrows. The lifetime of bubbles formed from nanodroplets ranged from a few transmissions/ milliseconds to dozens of transmissions/ milliseconds. If the imaging MI was low many vaporized droplets were not destroyed. The number of vaporization events was highest at the beginning of the acquisition and reduced over time.

\subsection{Contrast on-demand}
\begin{figure}[ht!]
  \centering
    \includegraphics[width=11cm]{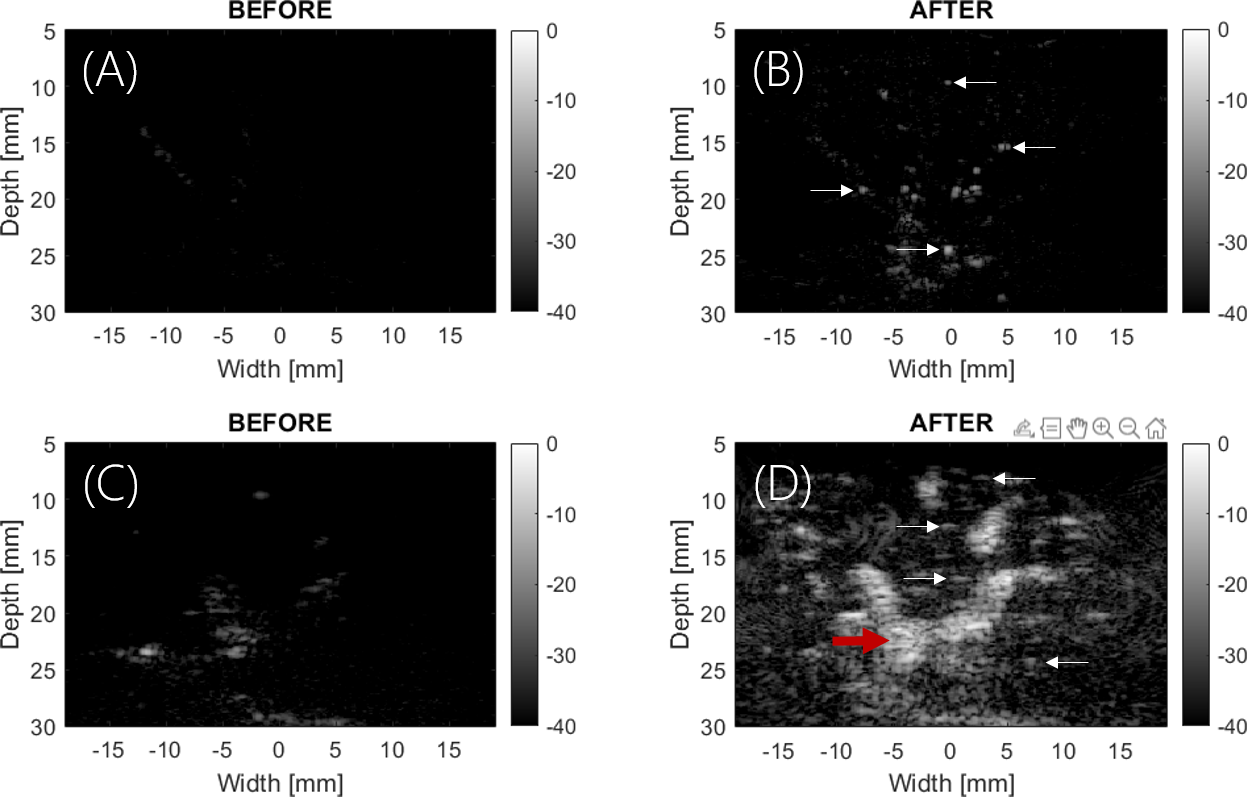}
      \caption{\textit{In vivo} imaging of rabbit kidney with nanodroplets. Filtered images before (A) and after (B) the first activation sequence of AWSALM (C4F10, single focus point at depth of 25 mm). The activation is focused on a single depth and bubbles are sparse. Filtered images before (C) and after (D) the first activation of fast-AWSALM (C3F8:C4F10). Activation occurs over the whole depth-range imaged. The white arrows in (B) and (D) show examples of isolated microbubbles.}
  \label{fig:03}
\end{figure}

The frame before and 15 frames after the first activation of C4F10 nanodroplets with AWSALM is shown in the clutter-filtered images in Figure \ref{fig:03} (A,B). It shows the local activation of C4F10 with a single focus point at 25 mm depth and subsequent occurrence of sparse bubbles in the medulla of a rabbit kidney. The white arrows indicate the location of exemplar microbubbles. The bubble density can be varied through the number of focus points and selection of the gas mixture. Figure \ref{fig:03} (C,D) demonstrates the activation of C3F8:C4F10 nanodroplets with the fast-AWSALM transmission sequence over a larger field of view than achieved with AWSALM. The distribution of microbubbles is denser in the medulla (large vessel) than in the cortex (small vessel) of the rabbit kidney. In the larger vessels, speckle patterns develop as indicated by the red arrow. In Figure \ref{fig:03} (C) a small number of bubbles can be observed prior to the activation pulse, likely due to spontaneous vaporization. The distribution and number of vaporization events is MI dependent (see supplementary Figure 3).

\subsection{\textit{In vivo} SR with AWSALM}
\begin{figure*}[ht!]
  \centering
    \includegraphics[width=15.5cm]{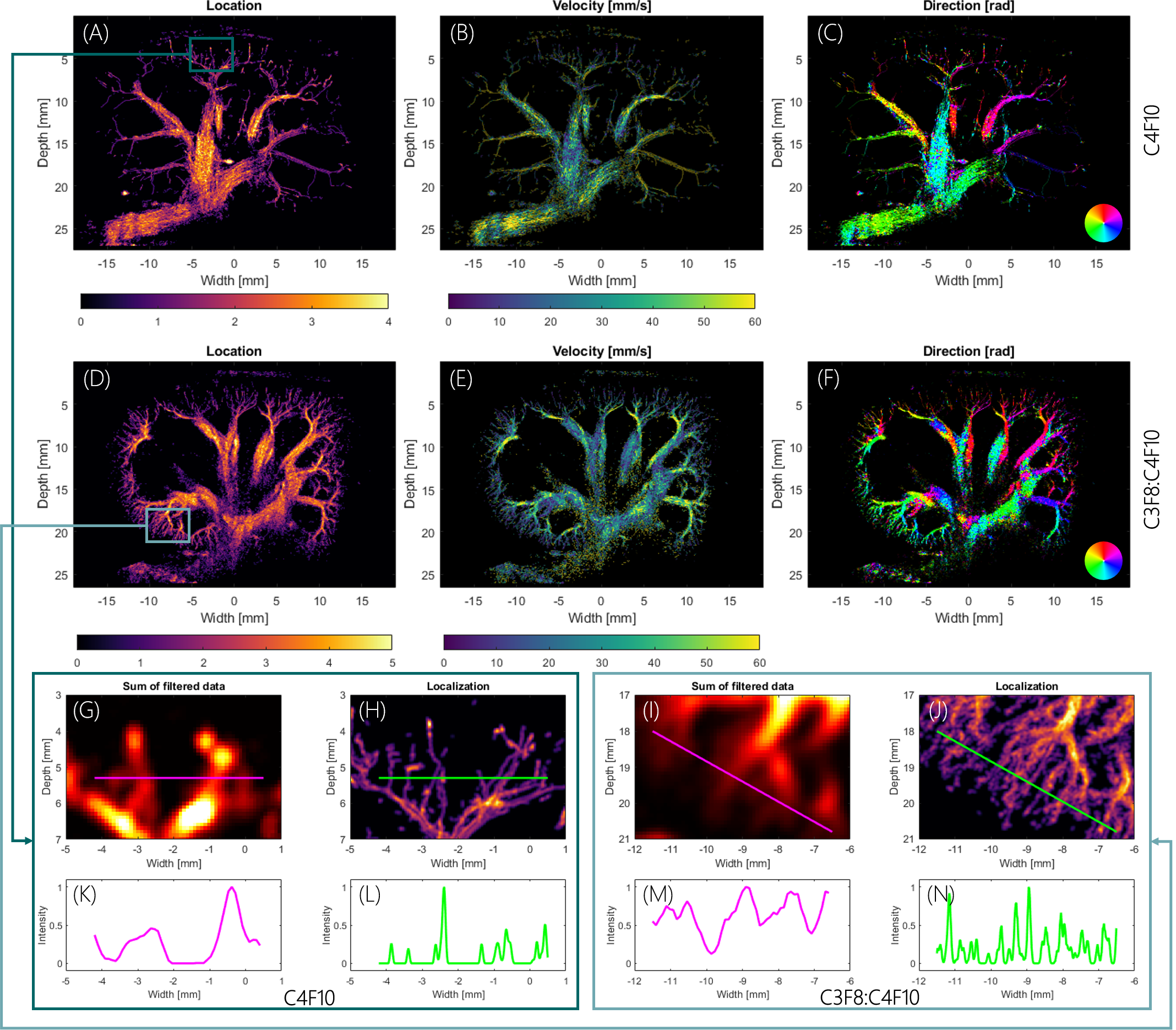}
      \caption{AWSALM super-resolution square rooted density map of localization (A,D), absolute flow velocity (B,E) and direction of droplet movement (C,F) in two rabbit kidneys. (A-C) presents a 3.5 s acquisition with C4F10 nanodroplets (3 s, imaging; 0.5 s activation). The focal region was at a depth of 25 mm and lateral positions from -4.8 mm to 2.4 mm. (D-F) show the maps of density, velocity and direction from an acquisition of C3F8:C4F10 droplets. Total duration of 5.5 s, of which 4.5 s and 1 s was used for imaging and activation respectively. Two focal zones at depths of 16 mm and 23 mm were used over a lateral range of -7.2 mm and 6.3 mm. (G-N) show a zoomed in view with corresponding normalized intensity profile.}
      \label{fig:05}
\end{figure*}

The AWSALM SR sequence was applied to two rabbit kidneys; the results (Figure \ref{fig:05}) demonstrate the versatility of the method. The top row maps the square rooted density of localized C4F10 droplets (A), their absolute velocity (B) and their direction (C). Data were acquired in 3.5 s, of which 3 s was imaging and 0.5 s was activation, respectively. Nanodroplets were activated at a focal depth of 25 mm, with 20 focused transmissions between -4.8 mm and 2.4. The bottom row shows a C3F8:C4F10 acquisition with 1 s of activation and 4.5 s of imaging. Activation occurred at two focal depths, the first at 16 mm and the second at 23 mm, with 20 focused transmissions between lateral positions of -7.2 mm and 6.3 mm. The peak velocities were 60 mm/s and both arterial blood flow and venous return could be quantified (see supplementary Video 2 and 3).

In Figure \ref{fig:05} (L,N) the mean vessel width at full width half maximum (FWHM) was 111 $\pm$ 39/ 102 $\pm$ 36 \textmu m for 9/ 16 vessels with an average of 1.84/ 3.81 number of localizations per vessel and a smallest measured vessel of 68 µm. Resolution was therefore increased by an 4-fold in the lateral and axial directions compared to the PSF at 4 MHz (543/ 530 \textmu m, see supplementary Figure 2). The smallest distance between two fully resolved adjacent vessels was 115 µm, which is smaller than half the wavelength (192.5 µm) at 4 MHz. Three times the number of vessels could be detected compared to conventional US.

\subsection{\textit{In vivo} SR with fast-AWSALM}
\begin{figure}[ht!]
  \centering
    \includegraphics[width=15cm]{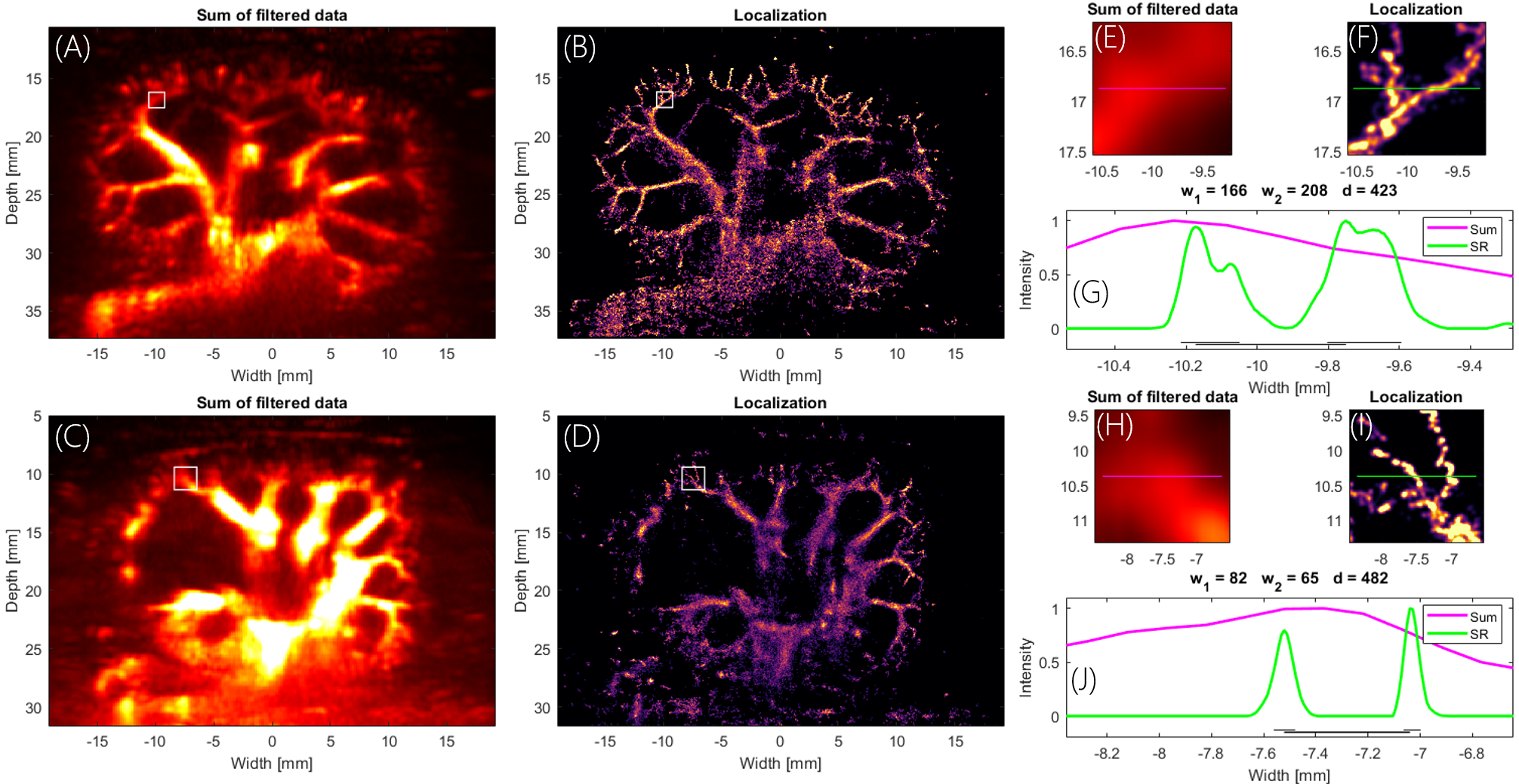}
      \caption{(A,C) Fast-AWSALM sum of filtered data, (B,D) SR density maps and (G,J) normalized intensity profile in region of interest shown by white boxes in (A-D). (E,F,H,I) zoomed in view of the region of interest. Images on the top (A,B,E-G) represent an acquisition with C3F8 nanodroplets (0.25 s, simultaneous imaging/activation). (C,D,H-J) show the results of imaging with the gas-mixture (C3F8:C4F10) droplets, obtained with 1 s of simultaneous imaging/activation. Data were acquired with a frame rate of 5,000 Hz and at MI=0.22.}
      \label{fig:06}
\end{figure}

Sub-second super-resolution imaging was achieved with the fast-AWSALM sequence and C3F8 droplets (Figure \ref{fig:06} A,B,E-G) and with C3F8:C4F10 droplets (Figure \ref{fig:06} C,D,H-J). Figure \ref{fig:06} (A,C,E,H) show the linear sum of the signal after filtering. Figure \ref{fig:06} (B,D,F,I) show the super-localized maps and Figure \ref{fig:06} (G,J) each show the respective normalized intensity profiles. The C3F8 droplet acquisition was acquired in 0.25 s and the C3F8:C4F10 acquisition was acquired in 1 s. For the C3F8:C4F10 mixture acquisition the majority of the signal was detected in larger vessels and vaporization events were mostly sparse and not connected in the cortex despite the longer duration. The signal from C3F8 droplets, on the other hand, was evenly distributed in the core and cortex region of the kidney. Both acquisitions reveal microvascular detail that is not visible without super-localized imaging (Figure \ref{fig:06} G,J). The accumulation of localization is shown in supplementary Video 2.

The smallest measured distance between two adjacent vessels was 423 µm (Figure \ref{fig:06}). The mean width was 130 $\pm 68$ \textmu m for 4 vessels and 2.8 number of localizations per vessel with the smallest measured vessel of 65 µm. This again yields a similar 4-fold improvement in the lateral and axial directions compared to the PSF at 4 MHz (487/ 574 \textmu m, see supplementary Figure 2).

\subsection{Selective super-localization}
\begin{figure}[ht]
  \centering
    \includegraphics[width=15cm]{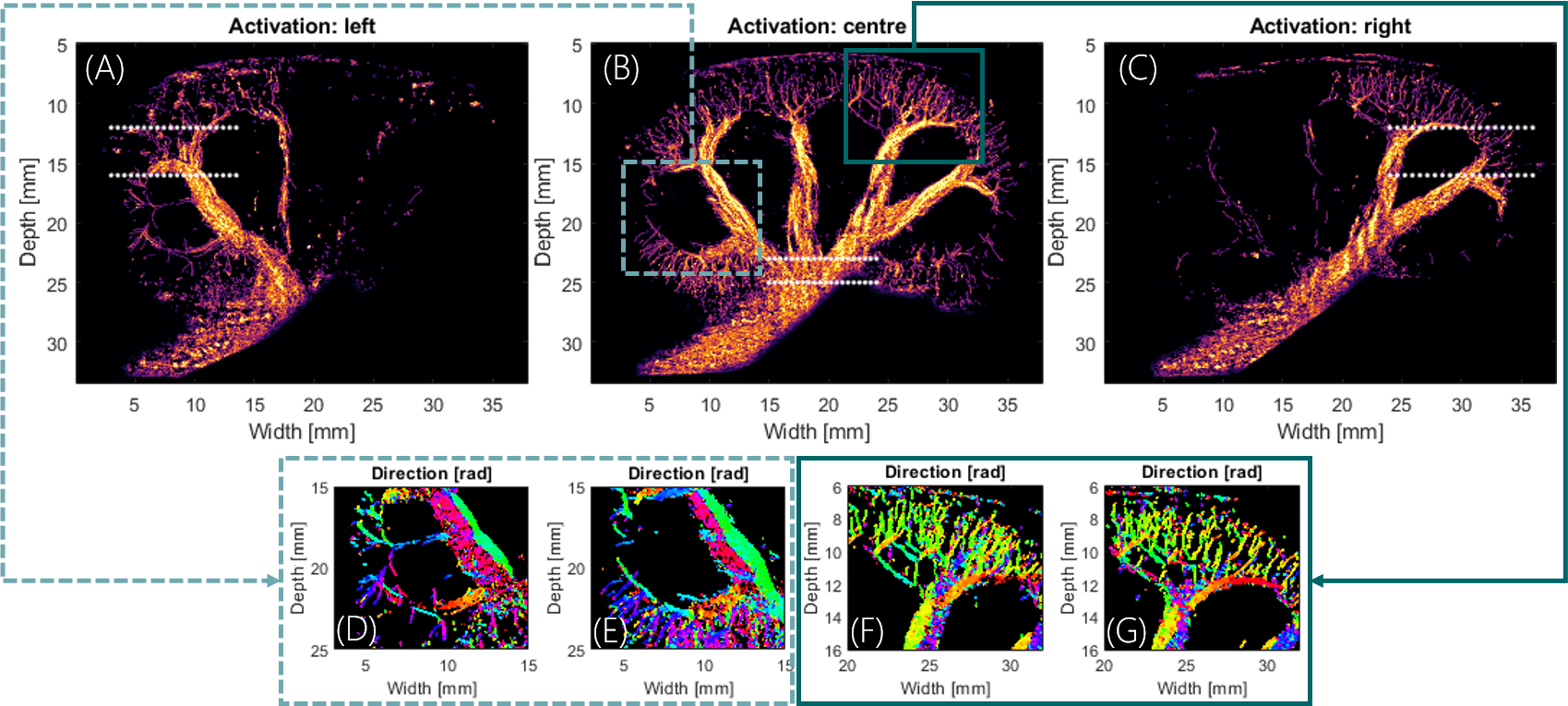}
      \caption{The deliberate activation and deactivation of nanodroplets with AWSALM can highlight different regions of the renal vasculature. (A-C) demonstrate the selective activation of different regions of the microvasculature in the same plane. (D,E) and (F,G) show flow direction in a region of interest, respectively.}
      \label{fig:07}
\end{figure}

The deliberate activation and deactivation of nanodroplets can be demonstrated through the selective activation of different regions of the microvasculature in the same imaging plane, as shown in Figure \ref{fig:07} (A-C). In Figure \ref{fig:07} (A) shows perfusion of the left branches, Figure \ref{fig:07} (B) perfusion of the entire renal vasculature and Figure \ref{fig:07} (C) structures on only the right side of the kidney. In Figure \ref{fig:07} (D-G), the direction of flow in a region of interest is shown. The dotted white lines in Figure \ref{fig:07} (A-C) show the focal points.

\subsection{Sub-second super-localization}
\begin{figure}[ht]
  \centering
    \includegraphics[width=15cm]{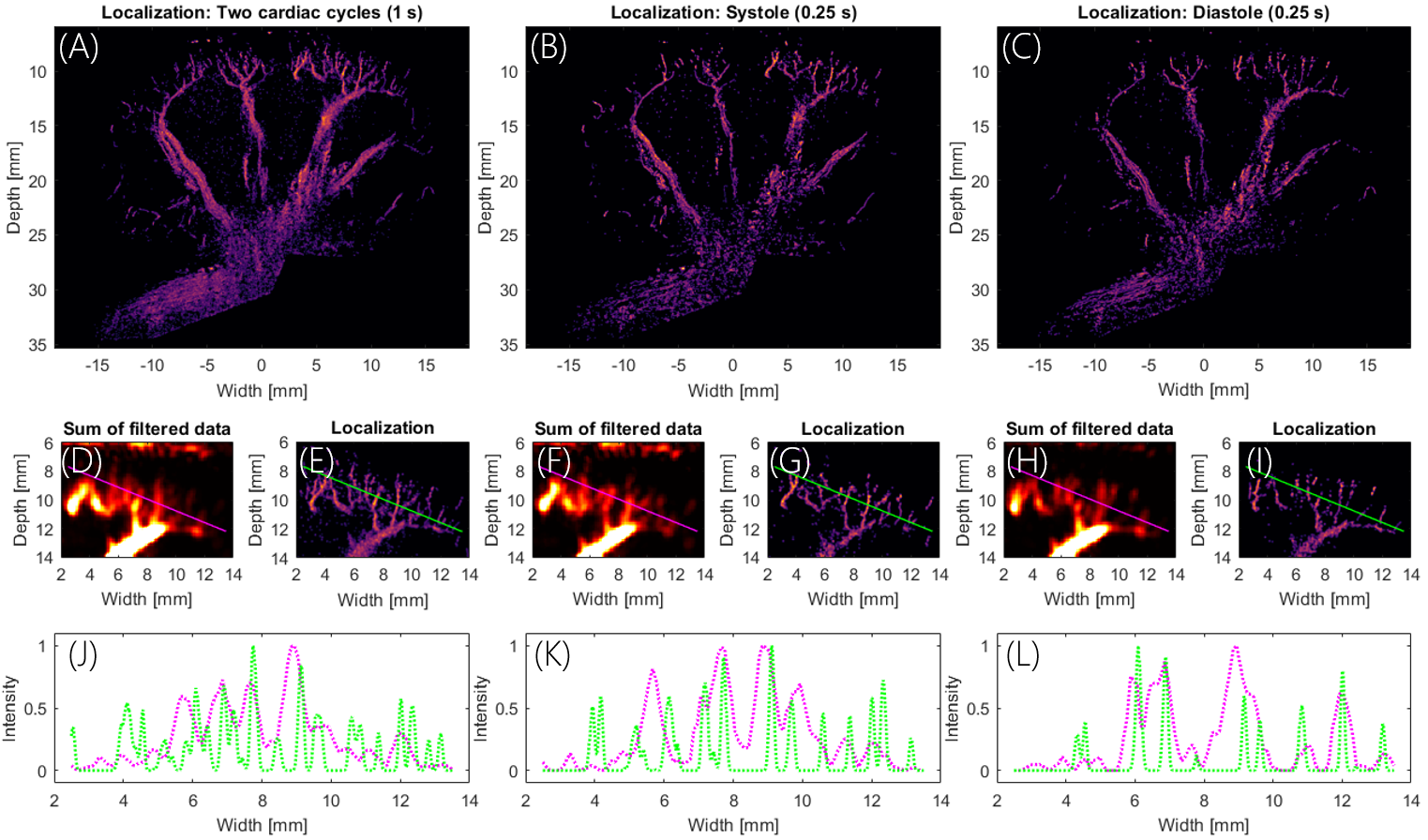}
      \caption{The activation of nanodroplets over the full field of view with fast-AWSALM can create sub-second visualizations of the renal vasculature, allowing separation of 0.25 second segments of systole (B) and diastole (C) from a single 1 s long acquisition (A). Different parts of the microvasculature are perfused with bubbles during the cardiac cycle. (D-L) show regions of interest with corresponding intensity profiles. Vessel number and location change between systolic, diastolic and two cardiac cycle perfusion.}
      \label{fig:08}
\end{figure}

The broad activation of nanodroplets with fast-AWSALM can be seen in Figure \ref{fig:08} (A-C), which show two sub-second visualizations of the rabbit kidney and microvascular structures during systolic (B) and diastolic (C) perfusion. Data were acquired with C3F8 droplets from the same 1 second acquisition (A). Each SR image was produced over a duration of 0.25 s; both were collected in the same cardiac cycle. In Figure \ref{fig:08} (D-L), a zoomed-in region of interest and corresponding normalized intensity profiles are shown. Some smaller vessels that are perfused during systole appear not to be perfused during diastole. The number of resolved adjacent vessels in the intensity profiles in (J,K,L) is 20, 13 and 10 respectively. For the localization over time see supplementary Video 3.

\section{Discussion}
The generation of ultrasound SR images poses two key challenges: how to obtain a sufficient contrast agent concentration to keep acquisition times low without compromising the ability to isolate individual microbubble signals; and how to actively manipulating contrast agent concentration in the blood stream in real time. In this study we demonstrated, for the first time, fast and selective super-localization imaging of microvascular flow \textit{in vivo} with AWSALM and fast-AWSALM using sono-switching nanodroplets. We demonstrated the capability of AWSALM to selectively visualize the downstream microcirculation of a specific vessel branch, and the feasibility of fast-AWSALM to capture the systolic and diastolic events in the kidney from sub-second acquisitions. Finally, we achieved this using acoustic pressures within recommended safety limits.

Sono-switchable nanodroplets enable generation of ultrasound contrast signals on demand, with control in both space and time (Figure \ref{fig:05} and \ref{fig:06}). By fast switching and imaging of these nanodroplets, AWSALM and fast-AWSALM offer the unique capability of generating both selective (Figure \ref{fig:07}) and fast (Figure \ref{fig:08}) super-localization microscopic images of the microcirculatory region of interest. At the same time, the techniques retain some key advantages of medical ultrasound, including deep tissue penetration and non-invasive imaging, when compared to optical techniques where penetration depth is an issue, or CT coronary angiography where contrast agents need to be directly injected into the artery of interest through a more invasive procedure. The microvascular contrast and achievable resolution also make the proposed techniques stand out among other clinical imaging modalities including CT, MRI and PET \cite{Dewey2020}.

The acoustic activation and deactivation of nanodroplets (Figure \ref{fig:03}, supplementary Video 1) is a unique feature that other non-invasive imaging modalities do not offer. It avoids the issues in perfusion imaging, that contrast agent kinetic modelling can suffer from effects of variable transit delays or fitting errors \cite{Krix2003}. Moreover, the selective activation of single branches allows direct measurement of blood velocity and vessel density or other key vascular performance indicators in the dependent vasculature. AWSALM SR could be used in the clinic for acute renal failure or transplant function, where contrast-enhanced ultrasound is already a widely applied method \cite{Krix2003, Wang2016a}.

Besides imaging the kidney as shown in this work, the deliberate activation of contrast agents with AWSALM could be applied to coronary angiography through minimal-invasive injection/ activation and super-resolution imaging of the coronary arteries in patients with coronary heart disease or arterial stenosis. Compared to x-ray or CT coronary angiography, AWSALM uses non-ionizing radiation and the contrast injection is less invasive. Ultrasound is already integrated into the clinical work flow and the application of AWSALM could visualise the functional status of the down stream microcirculation, e.g. how much blood flow there is within the specific micro-vessels downstream to the large coronary artery of interest. This could be highly valuable in clinical cardiology as studies have shown that a percentage of patients with coronary heart disease have microvascular disease while their large coronary arteries appears normal during angiographic assessment \cite{Aldiwani2021}. Such application could already be used in the clinic with condensation of clinically approved contrast agents.

AWSALM and fast-AWSALM can acquire images faster than microbubble-based ULM due to the higher concentration of contrast agents that can be injected at once. Moreover, the spatiotemporal features of nanodroplets (Figure \ref{fig:02}) facilitate their detection. In the spatiotemporal domain, activated droplets become lines with AWSALM and highly distinguishable short-lived dots with fast-AWSALM. These features could significantly speed up the detection algorithm. For example the slope of a bubble track in the spatiotemporal domain also corresponds to the axial or lateral velocity of the bubble.

We have shown in this study that the three different fluorocarbon nanodroplets can be vaporized in a microvascular flow mimicking phantom (Figure \ref{fig:02} and supplementary Video 1) and that the choice of droplet has a substantial impact on the SR imaging results (Figure \ref{fig:05}) affecting the number of vaporization events (which can also be controlled by adjusting the MI and frame rate). Through the use of the C3F8:C4F10 mixture the AWSALM sequence could be used at a significantly lower MI, while providing much greater stability and control for the fast-AWSALM sequence compared to C3F8 droplets. However, the main limitation of fast-AWSALM in the \textit{in vivo} experiments was variability in the amount of vaporization generated by same ultrasound transmission. This might be due to the rate of depletion of C3F8:C4F10 droplets. Changing the lipid shell \cite{Mountford2015a}, using mono-sized nanodroplets \cite{Seo2015} and implementing real time feedback to automatically adjust ultrasound parameters based on the received signals should be able to further improve the reproducibility of fast-AWSALM. Additionally, imaging MIs too high or too low do not generate satisfactory results. The former does not vaporize enough droplets, the latter destroys bubbles too fast.

When comparing AWSALM and fast-AWSALM, the latter is a closer counterpart of the PALM system for optical SR. AWSALM uses separate pulses for sono-switching and imaging. With this configuration it is possible to track the vaporized bubbles and generate flow velocity measurements. Fast-AWSALM uses the same plane wave pulses of medium MI for both sono-switching and imaging. With this configuration some vaporized bubbles are quickly destroyed, as seen in Figure \ref{fig:02}, making flow tracking more difficult (some bubbles are not destroyed). Image contrsat obtained with the current single-angle plane wave fast-AWSALM sequence might be improved (Figure \ref{fig:03}) with angle compounding (activated nanodroplets can sustain multiple transmissions Figure \ref{fig:02}).

In conclusion, this work shows the feasibility of on-demand and sub-second SR imaging with sono-switchable fluorocarbon nanodroplets for the selective real-time visualization of structural microvascular dynamics with AWSALM and fast-AWSALM. We demonstrated the feasibility of switching the nanodroplets at a relatively low MI, control of where (local or global) and when the switching takes place, flexibility over interleaving the switching pulses with imaging pulses, and the ability to change the vaporization threshold of the contrast agents. It is thus envisaged that the technique could be customized for a wide range of clinical or research needs.

\section{Acknowledgment}
This work was funded in part by the Engineering and Physical Sciences Research Council (EPSRC) under Grant EP/T008970/1, EP/T008067/1, and EP/N026942/1, the National Institute for Health Research i4i under Grant NIHR200972, and the British Heart Foundation under Grant PG/18/48/33832 .

\bibliographystyle{ieeetr}

\newpage
\section*{Supplementary material}

\begin{suppfigure}[ht]
  \centering
    \includegraphics[width=14cm]{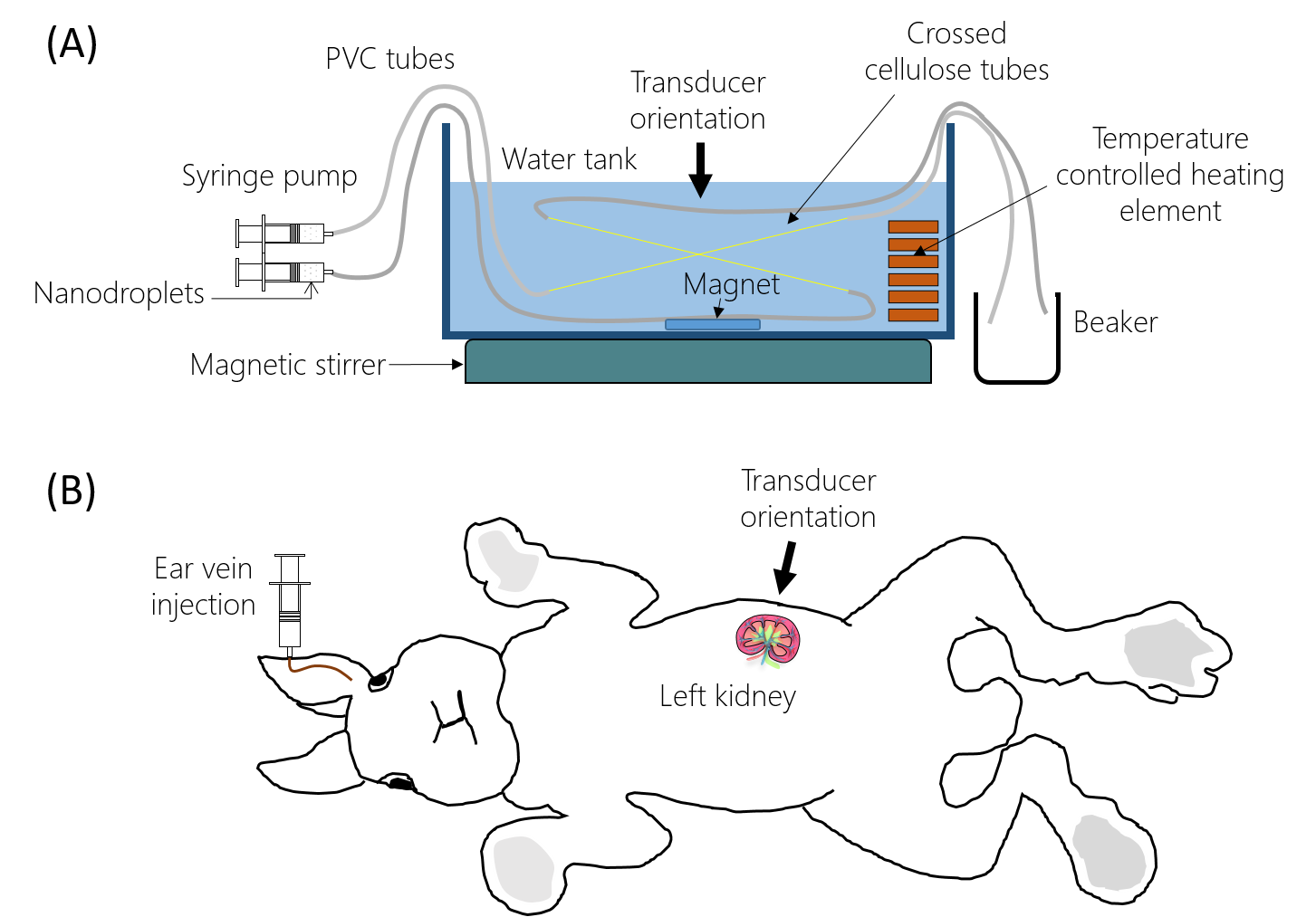}
      \caption{(A) In vitro crossed cellulose tube setup. The tube diameter is 200 \textmu m. (B) In vivo experimental setup and approximate transducer position indicated by arrow. Rabbits were shaved and laid flat on their back. The transducer was placed horizontally and slightly rotated to bring the renal artery into the field of view. The image is an own creation adapted from \cite{Zhou2018}}
      \label{supfig:00}
\end{suppfigure}

\begin{suppfigure}[ht]
  \centering
    \includegraphics[width=14cm]{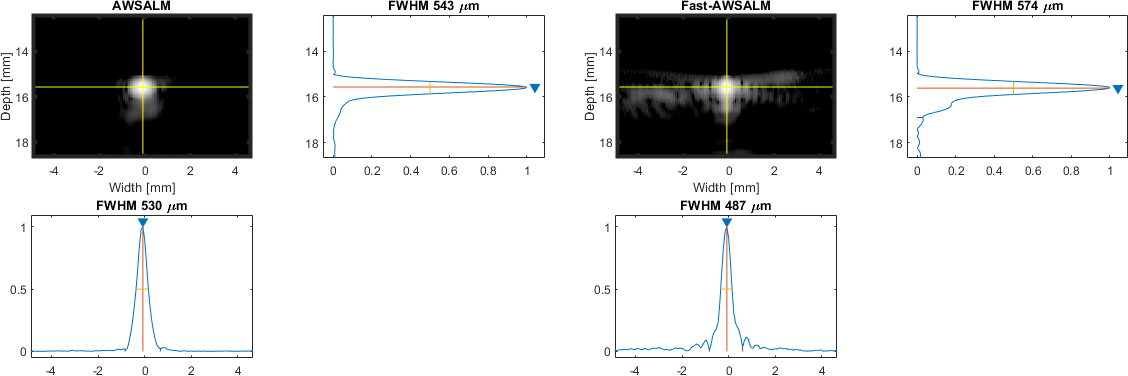}
      \caption{AWSALM and fast-AWSALM point spread function (PSF) calibrated with a 0.05 mm diameter stainless steel wire (Crazy Wire Company, UK) in the centre of the field of view at 15.6 mm depth and averaged over 100 frames.}
      \label{supfig:01}
\end{suppfigure}

\begin{suppfigure}[ht]
  \centering
    \includegraphics[width=14cm]{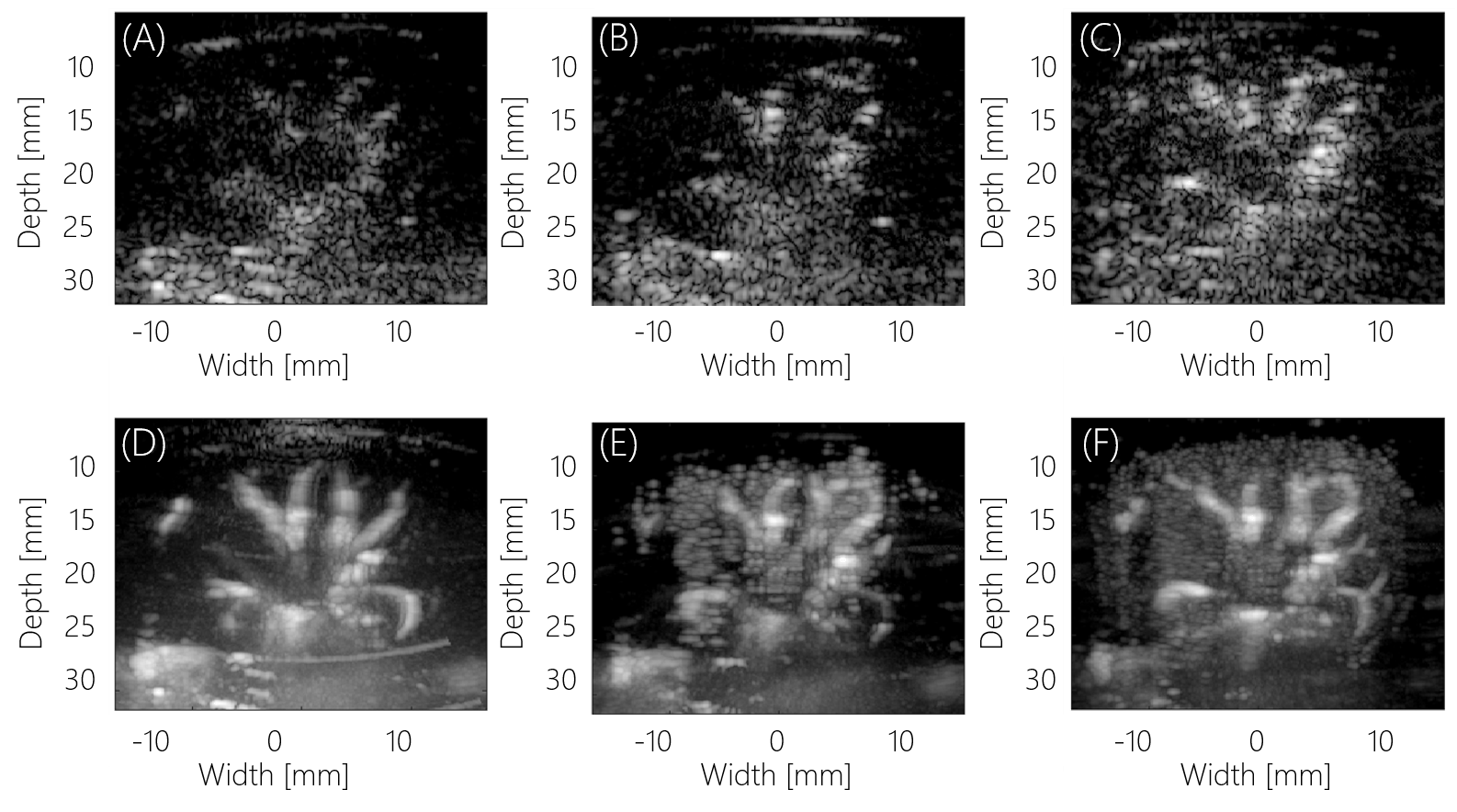}
      \caption{Mechanical index (MI) dependent activation of  C3F8:C410 nanodroplets in the renal vasculature of a rabbit. (A,B,C) show a single filtered frame and (D,E,F) show the maximum projection of 5000 frames acquired in 1 s. Between (A,B,C) the MI is increased from MI=0.1, MI=0.24 to MI=0.5. In line with the cross tube results, the number and spread of vaporization events with fast-AWSALM is MI dependent and for C3F8:C410 increases monotonically from MI=0.1 to MI=0.5 where it peaks. The spread of vaporization events subsequently reduces for higher MIs.}
      \label{supfig:02}
\end{suppfigure}

\end{document}